\documentclass[12pt]{article}
\usepackage{mathrsfs}
\usepackage{mathbbold}
\usepackage{amssymb}
\usepackage{epsfig}
\usepackage{graphicx}
\usepackage{amsmath}
\usepackage{verbatim}

\def\bit{\begin{itemize}} \def\eit{\end{itemize}}
\def\bec{\begin{center}}  \def\enc{\end{center}}
\def\be{\begin{equation}}
\def\ee{\end{equation}}
\def\ba{\begin{eqnarray}}
\def\ea{\end{eqnarray}}
\def\ben{\begin{eqnarray*}}
\def\een{\end{eqnarray*}}
\def\bec{\begin{center}}
\def\enc{\end{center}}
\def\bes{\begin{slide}}
\def\ens{\end{slide}}
\def\bei{\begin{itemize}}
\def\eei{\end{itemize}}

\renewcommand{\theequation}{\thesection.\arabic{equation}}
\csname @addtoreset\endcsname{equation}{section}

\begin{document}
\begin{titlepage}
\title{\bf\Large Nonlinear Realization of Spontaneously Broken $N=1$ Supersymmetry Revisited  \vspace{18pt}}

\author{\normalsize Hui Luo, Mingxing Luo and Liucheng Wang  \vspace{12pt}\\
{\it\small Zhejiang Institute of Modern Physics, Department of Physics,}\\
{\it\small Zhejiang University, Hangzhou 310027, P.R. China}\\
}

\date{}
\maketitle \voffset -.3in \vskip 1.cm \centerline{\bf Abstract}
\vskip .3cm
This paper revisits the nonlinear realization of spontaneously broken $N=1$ supersymmetry.
It is shown that the constrained superfield formalism as proposed in \cite{Seiberg}
can be reinterpreted in the language of standard realization of nonlinear supersymmetry via a new and simpler route.
Explicit formulas of actions are presented for general renormalizable theories with or without gauge interactions.
The nonlinear Wess-Zumino gauge is discussed and
relations are pointed out for different definitions of gauge fields.
In addition, a general procedure is provided to deal with theories of arbitrary Kahler potentials.

\vskip 5.cm \noindent January  2010
 \thispagestyle{empty}

\end{titlepage}
\newpage

\section{Introduction}

$N=1$ supersymmetry (SUSY) is arguably the most attractive extension of the standard model.
It provides a natural framework to include light scalars and is compatible with precision experiments.
It also provides tantalizing indirect evidences for grand unification theories.
Hopefully, it is to be discovered in LHC experiments.

If the fundamental theory is renormalizable, one may start with the linear realization of SUSY,
which can most conveniently be formulated by the notion of superspace (see, for example, \cite{WessBagger}).
Elements of the superspace are $(x,\theta,\bar\theta)$.
Superfields $\hat\Omega(x,\theta,\bar\theta)$ are functions of superspace\footnote
{To simplify presentation of the nonlinear theory, superfields and their components in the linear theory are hatted
while their counterparts in the nonlinear theory are not.}
with their component fields as coefficient functions of $x$ in their power series expansion in terms of $\theta$ and $\bar\theta$:
\be
\hat\Omega(x,\theta,\bar\theta) =\hat \phi + \theta \hat\psi_1 + \bar\theta \bar{\hat\psi}_2 + \theta \sigma^\mu \bar\theta \hat v_{\mu}
+ \theta^2 \hat F_1 + \bar\theta^2 \hat F_2
+ \theta^2 \bar\theta \bar{\hat\chi}_1 + \bar\theta^2 \theta \hat\chi_2 + \theta^2\bar\theta^2 \hat D
\ee
SUSY transformations on superfields are realized by differential operators $\xi Q + \bar\xi \bar Q$,
\be
\delta_\xi \hat\Omega(x,\theta,\bar\theta) = \left( \xi Q + \bar\xi \bar Q \right) \hat\Omega(x,\theta,\bar\theta) \label{lsusy}
\ee
which mix different components in $\hat\Omega(x,\theta,\bar\theta)$ linearly. Here
\be{\label{Qoperator}}
        Q_\alpha= \partial_\alpha -i (\sigma^\mu\bar{\theta})_\alpha \partial_\mu, \ \
        \bar{Q}_{\dot\alpha}=- \partial_{\dot\alpha} + i (\theta \sigma^\mu)_{\dot{\alpha}} \partial_\mu
\ee
Explicit transformation laws of component fields can be found from (\ref{lsusy}) by matching appropriate powers of $\theta$ and $\bar\theta$.
For many purposes, one defines two superspace covariant derivatives:
\be
        D_\alpha= \partial_\alpha + i (\sigma^\mu\bar{\theta})_\alpha \partial_\mu, \ \
        \bar{D}_{\dot\alpha}=- \partial_{\dot\alpha} - i (\theta \sigma^\mu)_{\dot{\alpha}} \partial_\mu
\ee
which anti-commutate with $Q_\alpha$ and $\bar{Q}_{\dot\alpha}$.

To be consistent with existing experiments,
the linear SUSY must be broken and broken spontaneously if one wishes to retain its salient virtues.
According to the general theory of spontaneously symmetry breaking,
there must exist a massless Goldstone fermion (Goldstino) field associated with this breaking.
The low energy physics related to Goldstino field could be relevant at the TeV scale\footnote
{Actually, the Goldstino becomes part of the massive gravitino when SUSY is spontaneously broken,
as (super-)gravity is omnipresent.
However, the lower energy physics will be dominated by the Goldstino,
if the SUSY breaking scale is not particularly high.}.
To deal with these physics,
it proves to be expedient to work with their nonlinearly realized versions.
It can be particularly useful if the system is strongly coupled, as exemplified by the low energy effective theory of hadronic physics \cite{Weinberg}.

Starting with the work of \cite{VA}, such a low energy effective theory has been developed over the years.
Most of the developments were summed in the framework of standard realization of nonlinear SUSY (see, for example, \cite{WessBagger}).
In this framework, a Goldstino field $\lambda$ is presumed to exist\footnote
{The construction of the $\lambda$ field out of fields in the linear theory had been discussed in \cite{IK3}.
For general O'Raifeataigh-like models, an explicit expression has been worked out in \cite{LLW}.
In this paper, this issue will not be addressed.}
which transforms as \cite{VA},
\be{\label{standard1}}
\delta_\xi \lambda_{\alpha}
=\frac{\xi_{\alpha}}{\kappa}- i \kappa (\lambda\sigma^\mu \bar{\xi} -\xi\sigma^\mu\bar\lambda) \partial_{\mu}\lambda_{\alpha}
\ee
under SUSY transformations.
Other fields are referred to as matter fields and they all are assumed to transform as,
\be
\delta_\xi \zeta=-i\kappa(\lambda\sigma^\mu \bar{\xi} -\xi\sigma^\mu\bar{\lambda}) \partial_{\mu} \zeta
{\label{standard2}}
\ee
To deal with chiral super-multiplets, it is convenient to define the chiral equivalents\footnote
{This is to be illustrated by discussions at the end of Section 3.1.
Notice that the roles of $(\tilde\lambda, \tilde\zeta)$ and $(\lambda, \zeta)$ are reversed from those in \cite{Wess83,LLW,LLZ},
to simplify presentations.
Also, the symbol $z$ is reserved for substitution rules presented in later sections.}
$\tilde\lambda$ and $\tilde\zeta$ via \cite{Wess83}
\be{\label{T}}
\tilde\lambda_\alpha(x)=\lambda_{\alpha}(w),\ \ \ \ \tilde\zeta(x) = \zeta(w),
\ee
with $w =x-i\kappa^{2} \lambda(w)\sigma \bar{\lambda}(w)$, such that the Goldstino and matter fields transform as,
\begin{equation}{\label{chiral1}}
\delta_{\xi}\tilde\lambda_{\alpha}
=\frac{\xi_{\alpha}}{\kappa}-2i\kappa\tilde\lambda\sigma^\mu \bar{\xi}\partial_\mu \tilde\lambda_{\alpha} , \ \ \
\delta_{\xi}\tilde\zeta=-2i\kappa\tilde\lambda\sigma^\mu \bar{\xi}\partial_\mu \tilde\zeta
\end{equation}

In some sense, it is a subject of ancient history, as much activities happened before the 1990s.
Renewed interests emerged recently.
In \cite{Seiberg}, a new approach has been proposed
by using constrained superfields instead of a manifestly nonlinear realization.
The Goldstino field is supposed to reside in a (constrained) chiral superfield $\hat X_{\rm NL}$.
The standard superspace technique is retained to write out Lagrangians while superfields are constrained to include only light degrees of freedom.
In \cite{LLZ}, it was proved that such a procedure can be reformulated in the language of standard realization of nonlinear SUSY.

Motivated by these developments, we revisit the nonlinear formulation of spontaneously broken $N=1$ linear SUSY.
In \cite{IK1,IK2,IK3}, it has been shown that any spontaneously broken linear SUSY theory
can be reformulated into a nonlinear one by appropriately changing superspace variables.
In this paper, this procedure will be adopted and converted to the notation of \cite{WessBagger}.
It will be shown that the constrained superfield formalism as proposed in \cite{Seiberg}
can be reinterpreted in the language of standard realization of nonlinear supersymmetry via a new and simpler route.
We will present explicit formulas of actions for all renormalizable theories with or without gauge interactions.
Particular attentions are paid to kinetic energies of chiral and vector superfields.
The nonlinear version of the Wess-Zumino gauge is discussed.
In addition, relations have been pointed out for canonical gauge fields and those emerged naturally in the nonlinear reformulation.
A general procedure is also provided to deal with theories of arbitrary Kahler potentials.

The paper is organized as follows.
Section 2 briefly reviews the general formalism of constructing nonlinear SUSY out of linear ones.
We apply the formalism to pure chiral superfields in Section 3
and to gauge superfields and gauge interactions in Section 4,
plus comparisons with the constrained superfield formalism.
We conclude in Section 5.
Listed in Appendix A are expressions for the Akulov-Volkov action as well as the
Jacobian of coordinate transformation.
Listed in Appendix B are Taylor expansions of nonlinear chiral superfields in terms of $\theta$ and $\bar\theta$.
Shown in Appendix C is a general procedure to deal with theories of arbitrary Kahler potentials.

\section{Constructing nonlinear SUSY out of linear SUSY, a brief review}

The general formalism in \cite{IK1,IK2, IK3} will be adopted, but in the notation of \cite{WessBagger}.
This section provides a brief review of the strategy, which will be elaborated more in later sections.
As shown in those papers, a set of nonlinearly realized fields can be obtained from a linear superfield
via a SUSY transformation parameterized by $\xi = -\kappa\lambda(x)$ and $\bar\xi = -\kappa\bar\lambda(x)$
\ba
              \Omega\left(x,\theta,\bar\theta\right) &=&
              \exp \left\{-\kappa\lambda(x) Q -\kappa \bar{\lambda}(x)\bar{Q}\right\} \hat\Omega\left(x,\theta,\bar\theta\right)
\ea
Under SUSY transformations,
\begin{eqnarray}\label{transform}
\delta_{\xi} \Omega =-i \kappa(\lambda\sigma^\mu \bar{\xi} -\xi\sigma^\mu\bar{\lambda})\partial_\mu \Omega
\end{eqnarray}
if $\hat\Omega$ transforms according to (\ref{lsusy}) and $\lambda$ according to (\ref{standard1}).
Defining new component fields in $\Omega$ according to its Taylor expansion in terms of $\theta$ and $\bar\theta$
\be
\Omega(x,\theta,\bar\theta) = \phi + \theta \psi_1 + \bar\theta \bar{ \psi}_2 + \theta \sigma^\mu \bar\theta   v_{\mu}
+ \theta^2 F_1 + \bar\theta^2 F_2
+ \theta^2 \bar\theta \bar{\ \chi}_1 + \bar\theta^2 \theta \chi_2 + \theta^2\bar\theta^2  D
\ee
we obtain $\phi, \psi_1, \cdots, D$ as composites of $\hat\phi, \hat\psi_1, \cdots,\hat D$, $\lambda$, and their spacetime derivatives.
According to (\ref{transform}), they all transform into themselves and independent of one another under SUSY transformations.
In particular, they satisfy the transformation law of (\ref{standard2}).

Observe that
\be
      \Omega\left(x,\theta,\bar\theta\right) = \hat\Omega\left(x +i\kappa\lambda(x)\sigma\bar{\theta}-i\kappa\theta\sigma\bar{\lambda}(x),
              \theta - \kappa\lambda(x),\bar \theta - \kappa\bar\lambda(x)\right)
\ee
That is, the nonlinear $\Omega$ can be obtained from the linear $\hat\Omega$
by replacing superspace arguments $(x, \theta, \bar{\theta})$ in the latter via
\be
\left\{ \begin{array}{lll}
  x &\rightarrow&  z=x +i\kappa\lambda(x)\sigma\bar{\theta}-i\kappa\theta\sigma\bar{\lambda}(x),  \\
  \theta &\rightarrow& \theta^{'} = \theta - \kappa\lambda(x),  \\
  \bar{\theta}  &\rightarrow& \bar{\theta}^{'} = \bar \theta - \kappa\bar\lambda(x).
\end{array} \right.\label{nc_sub}
\ee
Under this set of replacements, covariant derivatives $\partial_\mu, D_\alpha, \bar D_{\dot\alpha}$ are changed to
\be
\left\{ \begin{array}{lll}
\triangle_\mu&=& 
{(\emph{M}^{-1})_\mu}^\nu  \left(\nabla_\nu + \kappa \lambda_\nu^\alpha \partial_\alpha
+ \kappa \bar{\lambda}_{\nu\dot\alpha} \partial^{\dot\alpha} \right), \\
\triangle_\alpha & = & 
\partial_\alpha+i(\sigma^\mu \bar\theta)_\alpha\triangle_\mu, \\
\bar{\triangle}_{\dot\alpha}& = &
- \partial_{\dot\alpha} - i(\theta\sigma^\mu)_{\dot\alpha}\triangle_\mu,
\end{array} \right.
\ee
respectively.
Here $\nabla_\mu = {(\emph{T}^{-1})_\mu}^\nu (\partial/\partial x^\nu)$,
$\lambda_\mu = \nabla_\mu \lambda(x)$, $\bar\lambda_\mu = \nabla_\mu \bar\lambda(x)$, and
\be\left\{ \begin{array}{l}
{\emph{M}_\mu}^\nu(x,\theta,\bar{\theta})=\delta_{\mu}^{\nu}+i\kappa \lambda_{\mu}\sigma^{\nu}\bar{\theta}
                -i\kappa\theta\sigma^\nu \bar{\lambda}_\mu\\
{\emph{T}_\mu}^\nu(x) = \delta_\mu^\nu - i \kappa^2 \partial_\mu \lambda \sigma^\nu \bar \lambda + i \kappa^2 \lambda \sigma^\nu \partial_\mu \bar \lambda
\end{array} \right.\ee
This observation provides a simple procedure to convert any linear SUSY actions into their corresponding nonlinear ones.
For a generic action in the linear theory
\ba
S_{\rm gen}=\int d^4 x d^4\theta \ \mathcal{L}\left(\hat\Omega(x,\theta,\bar{\theta}),\partial_\mu \hat\Omega,
D_{\alpha}\hat\Omega,D_{\alpha} D_{\beta}\hat\Omega,...\right)
\ea
one replaces the integration variables $(x, \theta, \bar\theta)$ by $(z, \theta^{'}, \bar\theta^{'})$,
so $\hat\Omega(x,\theta,\bar{\theta}) \rightarrow \Omega(x,\theta,\bar{\theta})$ in $\mathcal{L}$
and the measure of integration becomes
\be
\int d^4 z d^4 \theta^{'}  = \int d^4 x d^4 \theta \det \emph{T}(x) \det \emph{M}(x,\theta,\bar\theta)
\ee
where $\det \emph{T} \det \emph{M}$ is the Jacobian.
Explicit expressions of $\det \emph{T}$ and $\det \emph{M}$ in terms of $\lambda$ are listed in Appendix A.
The end result is then
\ba
S_{\rm gen}^{\rm NL}=\int d^4 x d^4 \theta \det \emph{T} \det \emph{M} \
\mathcal{L}\left(\Omega(x,\theta, \bar\theta),\triangle_\mu\Omega,
\triangle_\alpha\Omega,\triangle_\alpha \triangle_\beta\Omega,...\right)
\ea
This completes the reformulation.

\section{Nonlinear reformulation of chiral superfields}
\renewcommand{\theequation}{\thesubsection.\arabic{equation}}
\csname @addtoreset\endcsname{equation}{subsection}
\subsection{Fields}

To deal with chiral/anti-chiral superfields, it is convenient to define the variable
$y = x + i \theta \sigma \bar\theta$ for chiral superfields and
its complex conjugate $y^\dagger = x - i \theta \sigma \bar\theta$ for anti-chiral superfields.
Covariant derivatives are then
\be
        D_\alpha^+= \partial_\alpha + 2i (\sigma^\mu\bar{\theta})_\alpha \partial_\mu, \ \
        \bar{D}_{\dot\alpha}^+=- \partial_{\dot\alpha}
\ee
in terms of $(y, \theta, \bar\theta)$ and
\be
        D_\alpha^-= \partial_\alpha , \ \
        \bar{D}_{\dot\alpha}^-=- \partial_{\dot\alpha} - 2 i (\theta \sigma^\mu)_{\dot{\alpha}} \partial_\mu
\ee
in term of $(y^\dagger, \theta, \bar\theta)$.
Replacing $(x, \theta,\bar{\theta})$ by $(z, \theta^{'},\bar{\theta}^{'})$ in (\ref{nc_sub}), one has
\be \begin{array}{l}
  y \rightarrow y - 2 i\kappa\theta\sigma\bar{\lambda}(x)+ i \kappa^2 \lambda(x) \sigma \bar\lambda(x) \\
  y^\dagger \rightarrow y^\dagger + 2 i\kappa\lambda(x)\sigma\bar{\theta} - i \kappa^2 \lambda(x) \sigma \bar\lambda(x)
\end{array}\ee
One is then led to define two sets of new substitution rules \cite{IK1,IK2,IK3},
\be
\left\{ \begin{array}{lll}
x &\rightarrow& z_+ =x - 2 i\kappa\theta\sigma\bar{\lambda}(x) + i \kappa^2 \lambda(x) \sigma \bar\lambda(x),
\ \ \ {\rm chiral} \\
 && z_- =x + 2 i\kappa\lambda(x)\sigma\bar{\theta} - i \kappa^2 \lambda(x) \sigma \bar\lambda(x),
  \ \ \ {\rm antichiral} \\
  \theta &\rightarrow&  \theta^{'} = \theta - \kappa\lambda(x), \\
  \bar{\theta}  &\rightarrow&  \bar{\theta}^{'} = \bar \theta - \kappa\bar\lambda(x)
\end{array} \right. \label{c_sub}
\ee
Accordingly, covariant derivatives $\partial_\mu, D_\alpha^\pm, \bar D_{\dot\alpha}^\pm$ are changed to
\be
\left\{ \begin{array}{l}
\triangle_\mu^\pm = {(\emph{M}^{-1}_\pm)_\mu}^\nu  \left(\nabla_\nu + \kappa \lambda_\nu^\alpha \partial_\alpha
+ \kappa  \bar{\lambda}_{\nu\dot\alpha}\partial^{\dot\alpha} \right) \\
\triangle_\alpha^+ = \partial_\alpha+2i(\sigma^\mu \bar\theta)_\alpha\triangle_\mu^+, \ \ \
\bar{\triangle}_{\dot\alpha}^+= - \partial_{\dot\alpha} \\
\triangle_\alpha^-  = \partial_\alpha,  \ \ \
\bar{\triangle}_{\dot\alpha}^-=
- \partial_{\dot\alpha}-2i( \theta\sigma^\mu )_{\dot\alpha}\triangle_\mu^-\\
{\emph{M}_{+\mu}}^\nu=\delta_{\mu}^{\nu} - 2 i\kappa\theta\sigma^\nu \bar{\lambda}_\mu, \ \ \
{\emph{M}_{-\mu}}^\nu = \delta_{\mu}^{\nu} + 2 i\kappa\lambda_\mu\sigma^\nu\bar\theta
\end{array} \right.
\ee

Linear chiral superfields are constrained by the condition $ \bar D_{\dot\alpha} \hat\Phi = 0$,
which can be solved by
$\hat\Phi(x, \theta, \bar\theta) = e^{i\theta \sigma^\mu \bar\theta \partial_\mu} \hat \varphi(x,\theta)$,
with $ \hat \varphi(x,\theta) = \hat\phi(x) + \sqrt{2} \theta \hat\psi(x) + \theta^2 \hat F(x)$.
A nonlinear $\Phi$ can be obtained from $\hat\Phi$ by replacing $(x, \theta,\bar{\theta})$ with $(z, \theta^{'},\bar{\theta}^{'})$ in the latter, \cite{IK1,IK2,IK3}
\be
\Phi(x, \theta, \bar\theta) = e^{i\theta \sigma^\mu \bar\theta \triangle_\mu^+}\varphi(x,\theta)
\ee
Here $ \varphi (x,\theta) = \phi + \sqrt{2} \theta \psi + \theta^2 F$,
obtained from $\hat \varphi(x,\theta)$ by $(x, \theta) \rightarrow (z_+, \theta^{'})$.
Similarly, one has from an anti-chiral superfield $\hat\Phi^\dagger(x, \theta, \bar\theta) = e^{-i\theta \sigma^\mu \bar\theta \partial_\mu} \hat \varphi^\dagger(x,\bar\theta)$,
\ba
\Phi^\dagger(x, \theta, \bar\theta) &=& e^{-i\theta \sigma^\mu \bar\theta \triangle_\mu^-} \varphi^\dagger(x,\bar\theta)
\ea
Here $\varphi^\dagger(x,\bar\theta)= \phi^\dagger + \sqrt 2 \bar \theta \bar \psi + \bar\theta^2 F^\dagger$,
obtained from $\hat \varphi^\dagger(x,\bar\theta)$ by $(x,  \bar\theta) \rightarrow (z_-, \bar\theta^{'})$ .
Taylor expansions of $\Phi(x, \theta, \bar\theta)$ and $\Phi^\dagger(x, \theta, \bar\theta)$
in terms of $\theta$ and $\bar\theta$ are listed in Appendix B.

In \cite{Seiberg}, a chiral superfield $\hat X_{\rm NL}(y)=\hat\phi_X+\sqrt 2\theta \hat\psi_X+\theta^2\hat F_X$
with the constraint $\hat X^2_{\rm NL}=0$ was proposed to describe the Goldstino field.
In this case, one can identify $\tilde\lambda$ with $\hat\psi_X / \sqrt 2\kappa\hat F_X$,
while its nonlinear version is $\varphi_{X}(x,\theta) =\theta^2 F_X$ \cite{LLW,LLZ}.
Clearly, $\varphi_{X}(x,\theta)$ is a projection operator,
which eliminates positive powers of $\theta$ in any nonlinear superfields via multiplication.
This observation provides a new and simpler route to reinterpret constraints in \cite{Seiberg}
in the language of standard realization of nonlinear SUSY,
as we shall see immediately and in Section 4.

To eliminate heavy components but to keep the light fermion component in $\hat Q_{\rm NL}(y)=\hat\phi_q+\sqrt 2\theta \hat\psi_q+\theta^2\hat F_q$,
the complex boson component in $\hat{\mathcal H}_{\rm NL}(y)=  \hat \phi_h + \sqrt2\theta \hat \psi_h + \theta^2 \hat F_h$,
and a real boson component in  $\hat{\mathcal A}_{\rm NL}(y)=  \hat \phi_a + \sqrt2\theta \hat \psi_a + \theta^2 \hat F_a$,
\cite{Seiberg} suggested to use constraints, $\hat X_{\rm NL} \hat Q_{\rm NL}=0$, $\hat X_{\rm NL} \bar{\hat {\cal H}}_{\rm NL}={\rm chiral}$,
and $\hat X_{\rm NL} (\hat {\mathcal A} - \bar{\hat {\mathcal A}}_{\rm NL})=0$, respectively.
Promoting these linear constraint equations into their nonlinear versions, one finds
\be \begin{array}{lll}
\hat X_{\rm NL} \hat Q_{\rm NL}=0 &\rightarrow& \theta^2 \varphi_Q(x,\theta) =0 \\
\hat X_{\rm NL} \bar{\hat {\cal H}}_{\rm NL}={\rm chiral} &\rightarrow& \theta^2 \varphi_{\cal H}^\dagger(x,\bar\theta) {\rm \ is \ independent \ of \ \bar\theta}  \\
\hat X_{\rm NL} (\hat {\mathcal A} - \bar{\hat {\mathcal A}}_{\rm NL})=0 &
\rightarrow& \theta^2 [ \varphi_{\cal A}(x,\theta)-\varphi_{\cal A}^\dagger(x,\bar\theta) ] = 0
\end{array}  \label{constraint qha}
\ee
respectively, since $\varphi_{X}(x,\theta)$ is proportional to $\theta^2$.
From (\ref{constraint qha}), one easily gets $\phi_q= \psi_h=F_h= \psi_a=F_a=0$, and $\phi_a=\phi^\dagger_a$, respectively.
In other words,
\be \begin{array}{ll}
\varphi_Q(x,\theta) = \sqrt 2\theta \psi_q+\theta^2 F_q, & \\
\varphi_{\cal H}(x,\theta) = \phi_h, & \\
\varphi_{\cal A}(x,\theta) = \phi_a, & {\rm here \ } \phi_a=\phi^\dagger_a
\end{array}
\ee
as one would have hoped.
These results can also be obtained by working directly with the component fields of $\hat Q_{\rm NL}$, $\hat {\cal H}_{\rm NL}$, or $\hat {\cal A}_{\rm NL}$.
However, such calculation could be formidable without a proper use of the chiral version Goldstino field $\tilde\lambda$.
Even with the help of $\tilde\lambda$, some of these calculations are still tedious and laborious.

\subsection{Actions}
Actions up to two spacetime derivatives for chiral superfields can generically be expressed as
the sum of a Kahler potential term $S_\mathcal{K}$ and a superpotential term $S_\mathcal{W}$
\be
S_{\rm ch} = \int d^4 x d^4\theta \ \mathcal{K}\left(\hat\Phi^\dagger(x,\theta,\bar{\theta}),\hat\Phi(x,\theta,\bar{\theta}) \right)
+\left\{  \int d^4 x d^2\theta \mathcal{W} \left( \hat\varphi(x, \theta)\right) +h.c. \right\}
\ee
For example, the most general renormalizable action for chiral superfields has the form
\ba
\mathcal{K}\left(\hat\Phi^\dagger, \hat\Phi \right) &=& \hat\Phi_i^\dagger \hat\Phi_i \nonumber \\
\mathcal{W}\left( \hat\Phi\right) &=& {1\over 2}m_{ij}\hat\Phi_i\hat\Phi_j +{1\over3}g_{ijk}\hat\Phi_i\hat\Phi_j\hat\Phi_k \nonumber
\ea
Here and hereafter, repeated indices $i, j$ are always summed over unless indicated otherwise.
The nonlinear version of the superpotential term $S_\mathcal{W}$ can be obtained by substituting integration variables
$(x, \theta, \bar\theta)$ with $(z_\pm, \theta^{'}, \bar\theta^{'})$,
such that $\hat\varphi(x, \theta)\rightarrow \varphi(x, \theta)$ in $\mathcal{W}$
and $\hat\varphi^\dagger(x, \bar\theta)\rightarrow \varphi^\dagger(x, \bar\theta)$ in $\mathcal{W}^\dagger$, respectively.
The measures are then
\be
\left. \begin{array}{l}
\int d^4 z_+ d^2 \theta^{'} = \int d^4 x d^2 \theta  \det \emph{T} \det \emph{M}_+ , \\
\int d^4 z_- d^2 \bar\theta^{'} = \int d^4 x d^2 \bar\theta  \det \emph{T} \det \emph{M}_-,
\end{array}\right.
\ee
of Jacobians
\be\begin{array}{l}
\det \emph{M}_{+} = 1 - 2 i\kappa\theta\sigma^\mu\bar{\lambda}_\mu
+ 4 \kappa^2\theta^2 \bar{\lambda}_\mu \bar\sigma^{\nu\mu} \bar{\lambda}_\nu,\\
\det \emph{M}_- =1 + 2 i\kappa\lambda_\mu\sigma^\mu\bar\theta
+ 4 \kappa^2 \bar\theta^2 \lambda_\mu \sigma^{\nu\mu} \lambda_\nu.
\end{array}
\ee
respectively. The end result is
\be
S_{\mathcal{W}}^{\rm NL}=\int d^4 x d^2 \theta \det \emph{T} \det \emph{M}_+ \
\mathcal{W}\left(\varphi(x,\theta)\right) +h.c.
\ee

Grassmann integrations in $S_{\mathcal{W}}^{\rm NL}$ can be easily worked out by noting that
\be
\mathcal{W}(\varphi) = \mathcal{W}(\phi) + \sqrt2  {\partial \mathcal{W}(\phi) \over \partial \phi_i} \theta \psi_i
+ \theta^2 \left( {\partial \mathcal{W}(\phi) \over \partial \phi_i} F_i
- {1\over 2} {\partial^2 \mathcal{W}(\phi) \over \partial \phi_i \partial \phi_j} \psi_i \psi_j \right) \nonumber
\ee
So $S_{\mathcal{W}}^{\rm NL}$ is the real part of
\be
 2 \int d^4 x \det \emph{T}
 \left\{ {\partial \mathcal{W} \over \partial \phi_i} \left(  F_i +  \sqrt2 i \kappa\psi_i\sigma^\mu\bar{\lambda}_\mu \right)
 - {1\over 2} {\partial^2 \mathcal{W} \over \partial \phi_i \partial \phi_j} \psi_i \psi_j
 + 4 \kappa^2 \mathcal{W} \bar{\lambda}_\mu \bar\sigma^{\nu\mu} \bar{\lambda}_\nu \right\}
\ee
The Kahler potential term can be dealt with in the same manner of $S_{\rm gen}$
\be
S_{\mathcal{K}}^{\rm NL}=\int d^4 x d^4 \theta \det \emph{T} \det \emph{M} \
\mathcal{K}\left(\Phi^\dagger(x,\theta,\bar{\theta}),\Phi(x,\theta,\bar{\theta}) \right)
\ee
But this form is complicated and will be systematically treated in Appendix C. Significant simplification can be achieved by noticing that
\be
S_{\mathcal{K}} = {1\over 2} \int d^4 x d^4\theta
\left\{ \mathcal{K}\left( e^{-2i\theta \sigma^\mu \bar\theta \partial_\mu} \hat \varphi^\dagger, \hat \varphi \right)
+ \mathcal{K}\left( \hat \varphi^\dagger, e^{2i\theta \sigma^\mu \bar\theta \partial_\mu} \hat \varphi \right) \right\}
\ee
up to surface terms which do not affect perturbation theories. The symmetric form is to ensure $S_{\mathcal{K}} $ to be Hermitian.
Note that
\be\left\{\begin{array}{l}
\hat \varphi(x, \theta) \rightarrow \varphi(x, \theta)  \\
e^{-2i\theta \sigma^\mu \bar\theta \partial_\mu} \hat \varphi^\dagger(x,\bar{\theta}) \rightarrow
e^{-2i\theta \sigma^\mu \bar\theta \triangle_\mu^-} \varphi^\dagger(x,\bar\theta)
\end{array}\right.
\ee
if $(x, \theta, \bar\theta)$ are replaced by $(z_+, \theta^{'}, \bar\theta^{'})$, and
\be\left\{\begin{array}{l}
e^{2i\theta \sigma^\mu \bar\theta \partial_\mu} \hat \varphi(x, \theta) \rightarrow
e^{2i\theta \sigma^\mu \bar\theta \triangle_\mu^+} \varphi(x, \theta)  \\
\hat \varphi^\dagger(x,\bar{\theta}) \rightarrow
 \varphi^\dagger(x,\bar\theta)
\end{array}\right.
\ee
if $(x, \theta, \bar\theta)$ are replaced by $(z_-, \theta^{'}, \bar\theta^{'})$.
Thus, $S_{\mathcal{K}}$ can be treated in the same manner as $S_{\mathcal{W}}$ and one has
\be
S_{\mathcal{K}}^{\rm NL}=
\int d^4 x d^4\theta {\det \emph{T}\over 2}  \left\{\det \emph{M}_+
\mathcal{K}\left( e^{-2i\theta \sigma^\mu \bar\theta \triangle_\mu^-}  \varphi^\dagger, \varphi \right)
+\det \emph{M}_- \mathcal{K}\left(  \varphi^\dagger, e^{2i\theta \sigma^\mu \bar\theta \triangle_\mu^+} \varphi \right) \right\}
\ee

For the canonical Kahler potential $\hat\Phi^\dagger \hat\Phi$, we have
\ba
S_{\mathcal{K}}^{\rm NL}&=& \Re \int d^4x \det\emph{T} \left\{ \phi\nabla^\mu \phi_\mu^\dagger-i\psi\sigma^\mu\nabla_\mu\bar\psi+F F^\dagger \right.\\
&&+\sqrt 2 \kappa \left( 2\phi \bar\lambda_\nu\bar\sigma^{\nu\mu}\nabla_\mu\bar\psi
+\phi^\dagger_\mu \lambda_\nu\sigma^\mu\bar\sigma^\nu\psi\right)
 \left. -4 i \kappa^2 \phi \phi^\dagger_\mu \lambda_\nu\sigma^\mu\bar\sigma^{\nu\rho}\bar\lambda_\rho \right\}\nonumber
\ea
with $\phi_\mu^\dagger$ given in (\ref{phipsimu}).

As in the case of linear SUSY, the $F$ field is quadratic in $S_{\rm ch}^{\rm NL}$
so it can be integrated out via its equation of motion.
For the particular Kahler potential here, one actually has the same form of that in the linear theory,
\be
F = -\left( {\partial \mathcal{W} \over \partial \phi} \right)^\dagger  \label{f_eom}
\ee

\section{Nonlinear reformulation of vector superfields}

\subsection{Fields}

Vector superfields are usually associated with the SUSY generalization of gauge transformation.
They are constrained by the condition
\be
\hat V(x,\theta,\bar\theta) = \hat V^\dagger (x,\theta,\bar\theta)
\ee
The nonlinear version $V$ can be obtained from $\hat V$ by substitution rules in (\ref{nc_sub}).
Due to gauge symmetries, either $\hat V$ or $V$ can always be put into the so-called Wess-Zumino gauge, but not simultaneously.
In this paper, $V$ will be chosen in the Wess-Zumino gauge to simplify the presentation of nonlinear theory, in the same manner as \cite{IK3}.
In \cite{LLZ}, this was proven to be equivalent to the constraint $\hat X_{\rm NL} \hat V=0$ as proposed in \cite{Seiberg},
by working directly with component fields of $\hat V$.
However, this equivalence can be more easily proved by promoting the constraint equation into its nonlinear version and making use of the fact $X_{\rm NL} \sim \theta^2$ again.
Specifically,
\ba
\hat X_{\rm NL} \hat V=0 &\rightarrow& \theta^2 V = 0
\ea
so the coefficient functions of $1$, $\bar\theta$, and $\bar\theta^2$ all vanish.
Since $V$ is a real superfield, the coefficient functions of $\theta$ and $\theta^2$ also vanish.
This yields nothing but the nonlinear Wess-Zumino gauge condition on $V$.
For later convenience, we will choose
\ba
V &=& - \theta \sigma^\mu \bar \theta v_\mu + i \theta^2 \bar
\theta (\bar\chi-{1\over 2} \kappa \bar\sigma^\mu \sigma^\nu
\bar\lambda_\mu v_\nu )
- i \bar\theta^2 \theta (\chi-{1\over 2}\kappa\sigma^\mu \bar\sigma^\nu \lambda_\mu v_\nu)\nonumber\\
&& + {1\over 2} \theta^2\bar\theta^2 (D-\kappa^2 \lambda_\mu \sigma^\nu \bar \lambda^\mu v_\nu)
\ea
where $v_\mu$ and $D$ are real fields.
The unconventional choices of $\chi$ and $D$ are to ensure
that they have simple transformation properties in the residual gauge symmetry within the Wess-Zumino gauge,
as to be seen below.

For simplicity, we start with Abelian gauge theories\footnote
{Extension to non-Abelian gauge theories is straightforward, as to be shown below.}.
The linear supersymmetric field strength is then $\hat W_\alpha = -{1\over 4}\bar D^2 D_\alpha \hat V$.
Since $\hat W_\alpha$ is a chiral superfield, it can be most conveniently evaluated in the form of
\be
\hat W_\alpha = -{1\over 4} e^{i \theta \sigma^\mu \bar \theta \partial_\mu} (\bar D^+)^2 D_\alpha^+ \hat U
\ee
here $\hat U$ is related to $\hat V$ via
\be\label{relation}
\hat U(x, \theta, \bar\theta) = \hat V(x-i \theta \sigma \bar \theta, \theta, \bar\theta)
\ee
Promote $\hat V$ into $V$ via $(x, \theta, \bar\theta) \rightarrow (z, \theta^{'}, \bar\theta^{'})$
and $\hat U$ into $U$  via $(x, \theta, \bar\theta) \rightarrow (z_+, \theta^{'}, \bar\theta^{'})$
\ba
V&=&\hat V(x+i\kappa \lambda \sigma \bar \theta -i\kappa \theta \sigma \bar \lambda,\theta-\kappa \lambda,\bar\theta-\kappa\bar\lambda)\\
U&=&\hat U(x-2 i \kappa \theta \sigma \bar \lambda+ i \kappa^2 \lambda \sigma \bar \lambda,\theta- \kappa \lambda,\bar\theta-\kappa\bar\lambda) \nonumber\\
&=&\hat V(x+i\kappa \lambda \sigma \bar \theta -i\kappa \theta \sigma \bar \lambda-i\theta \sigma \bar\theta,\theta-\kappa \lambda,\bar\theta-\kappa\bar\lambda)
\ea
from which one has $U = \exp(-i\theta \sigma^{\mu} \bar{\theta}\triangle_\mu) V$. In components,
\ba
U&=&-\theta \sigma^\mu \bar \theta v_\mu + i \theta^2 \bar \theta \bar\chi
-i \bar\theta^2 \theta (\chi-\kappa\sigma^\mu \bar\sigma^\nu \lambda_\mu v_\nu)\nonumber\\
&&+ \frac{1}{2}\theta^2\bar\theta^2 (D-i \nabla_\mu v^\mu
+\kappa \lambda_\mu \sigma^\mu \bar \chi-\kappa\chi\sigma^\mu \bar\lambda_\mu
- 2\kappa^2 \lambda_\mu \sigma^\rho  \bar \lambda^\mu v_\rho)
\ea
In \cite{IK3}, the combination $D+\kappa \lambda_\mu \sigma^\mu \bar \chi-\kappa\chi\sigma^\mu \bar\lambda_\mu$ was identified as the $D$ field,
which turns out to be complex and inconvenient.

The nonlinear supersymmetric field strength is then
\ba
W_\alpha &=& -{1\over 4} e^{i\theta \sigma^{\mu} \bar{\theta}\triangle_\mu^+}
 \bar \triangle^+_{\dot\alpha} \bar \triangle^{+\dot\alpha} \triangle^+_\alpha U
 =  e^{i\theta \sigma^{\mu} \bar{\theta}\triangle_\mu^+} \varpi_\alpha
\\
\varpi_\alpha&=&- i \chi_\alpha + \theta_\alpha D-{i\over 2} (\sigma^\mu \bar \sigma^\nu \theta)_\alpha F_{\mu\nu}
- 2 \kappa (\sigma^{\mu\nu} \theta)_\alpha (\chi \sigma_\nu \bar\lambda_\mu - \lambda_\mu \sigma_\nu \bar\chi)
 \label{W_NL}\\
&& + \theta^2 \left[ (\sigma^\mu \nabla_\mu\bar \chi)_\alpha - i\kappa (\sigma^\mu \bar{\lambda}_\mu)_\alpha D
   - {\kappa\over 2} (\sigma^\mu \bar \sigma^\gamma \sigma^\nu \bar\lambda_\mu)_\alpha F_{\nu\gamma}
   \right.\nonumber \\
&&
 \left. +2i \kappa^2 (\sigma^\mu \bar\sigma^{\rho\nu} \bar{\lambda}_\mu)_\alpha
 (\chi \sigma_\rho \bar\lambda_\nu - \lambda_\nu \sigma_\rho \bar \chi) \right] \nonumber
\ea
where
\ba
F_{\mu\nu} &=&\nabla_\mu v_{\nu} - \nabla_\nu v_{\mu}
+2 i \kappa^2 \left(\lambda_\mu \sigma^\rho \bar \lambda_\nu -  \lambda_\nu \sigma^\rho \bar \lambda_\mu \right) v_{\rho}
\ea
Notice that $W_\alpha$ is invariant under the following residual gauge transformation \cite{IK3}:
\ba
\delta_\alpha v_\mu= \nabla_\mu \alpha(x),\ \ \ \delta_\alpha \chi = 0,\ \ \ \delta_\alpha D = 0 \label{r_gauge}
\ea
One sees that both $\chi$ and $D$ are invariant under this transformation,
which supports the rational for choosing the particular form of $V$.

In formulations starting with nonlinear fields directly (see, for example, \cite{CL04,CL96,Klein,LLZ}),
one usually uses the canonical gauge field $v_\mu^C$ instead of the $v_\mu$ defined here.
Under gauge rotations, $v_\mu^C$ transforms as
\be
\delta_\alpha v^{C}_\mu = \partial_\mu \alpha
\ee
They are related to each other simply via
\be
v^{C}_\mu  = T_\mu^\nu v_\nu
\ee
Under SUSY transformations,
\ba
\delta_\xi v_\mu &=& - i\kappa (\lambda\sigma^\nu \bar\xi - \xi \sigma^\nu \bar \lambda) \partial_\nu v_\mu
\\
\delta_\xi v^{C}_\mu &=& - i\kappa (\lambda\sigma^\nu \bar\xi - \xi \sigma^\nu \bar \lambda) \partial_\nu v^{C}_{\mu}
- i \kappa \partial_\mu (\lambda\sigma^\nu \bar\xi - \xi \sigma^\nu \bar \lambda) v^{C}_\nu
\ea
The canonical field strength is defined as
\be
F^{C}_{\mu\nu} = (T^{-1})_\mu^\rho (T^{-1})_\nu^\sigma (\partial_\rho v^{C}_\sigma - \partial_\sigma v^{C}_\rho)
\ee
such that $F_{\mu\nu}=F^{C}_{\mu\nu}$.
They transform co-variantly under gauge transformations
and as matter fields under non-linear SUSY transformations.

For non-Abelian gauge fields,
the nonlinear supersymmetric field strength is
\be
W_\alpha = -{1\over 8g} e^{i\theta \sigma^{\mu} \bar{\theta}\triangle_\mu^+}
 \bar \triangle^+_{\dot\alpha} \bar \triangle^{+\dot\alpha}  e^{-2gU} \triangle^+_\alpha e^{2gU}
\ee
where $W_\alpha = W_\alpha^a T^a$ and $U = U^a T^a$.
$W_\alpha$ can be obtained from (\ref{W_NL}) by simple modifications:
\be\left\{\begin{array}{l}
F_{\mu\nu} \rightarrow \nabla_\mu v_{\nu} - \nabla_\nu v_{\mu}
+2 i \kappa^2 \left(\lambda_\mu \sigma^\rho \bar \lambda_\nu -
\lambda_\nu \sigma^\rho \bar \lambda_\mu \right) v_{\rho} + i g [v_\mu, v_\nu], \\
\nabla_\mu \chi \rightarrow \nabla_\mu \chi + i g [v_\mu, \chi]  \\
v_\mu = v_\mu^a T^a,\ \ \ \chi = \chi^a T^a, \ \ \ D=D^a T^a
\end{array}\right. \label{non-abelian}
\ee
where $T^a$ is the adjoint representation matrix of the gauge group.
$W_\alpha$ is covariant under the following residual gauge transformation:
\ba
\delta_\alpha v_\mu = \nabla_\mu \alpha(x) + i [ \alpha(x), v_\mu],\ \ \
\delta_\alpha \chi = i [\alpha(x), \chi],\ \ \ \delta_\alpha D = i [\alpha(x), D]
\ea
Note that the gauge group can be arbitrary and need not to be simple.

\subsection{Actions}
We start with the Abelian gauge group and extensions to arbitrary gauge groups are straightforward.
The kinetic energy for an Abelian gauge field up to two spacetime derivatives is given by
\be
S_{\rm V}={1\over 4} \int d^4 x \left( \int d^2 \theta \mathscr{H}(\hat\Phi) \hat W^{\alpha} \hat W_{\alpha}+h.c\right)
\ee
which can be treated in the same ways as $S_{\mathcal W}$ by replacing $(x, \theta, \bar\theta)$ with $(z_\pm, \theta^{'}, \bar\theta^{'})$ to get
\be
S_{\rm V}^{\rm NL}= {1\over 4} \int d^4 x \det \emph{T} \left( d^2\theta  \det \emph{M}_+ \mathscr{H}(\varphi) \varpi^{\alpha} \varpi_{\alpha} + h.c. \right)
\ee
Integrating out the Grassmann variables, terms inside the parentheses become
\ba
\mathscr{H}(\phi)  F _{\varpi \varpi } - \left( {\partial \mathscr{H}(\phi) \over \partial \phi_i} F_i
- {1\over 2} {\partial^2 \mathscr{H}(\phi) \over \partial \phi_i \partial \phi_j} \psi_i \psi_j \right) \chi^{2}
- 4 \kappa^2  \bar{\lambda}_\mu \bar\sigma^{\nu\mu} \bar{\lambda}_\nu \mathscr{H}(\phi) \chi^{2} \\
- {1\over \sqrt2} {\partial \mathscr{H}(\phi) \over \partial \phi_i} \psi_i \psi_{\varpi \varpi }
- \sqrt2 i \kappa {\partial \mathscr{H}(\phi) \over \partial \phi_i} \psi_i \sigma^\mu\bar{\lambda}_\mu \chi^{2}
+ i \kappa  \mathscr{H}(\phi)\psi_{\varpi \varpi } \sigma^\mu\bar{\lambda}_\mu +h.c. \nonumber
\ea
where the $\varpi\varpi$ indexed fields are from
$\varpi^\alpha \varpi_\alpha = -\chi^{2} +  \theta \psi_{\varpi \varpi } +  \theta^2 F_{\varpi \varpi }$, in which
\ba
\psi^{\varpi \varpi}_\alpha  = & \hspace{-5pt}-2iD\chi_\alpha-2i\kappa\lambda_{\mu}\sigma^{\mu}\bar{\chi}\chi_\alpha
-3i\kappa\chi^{2}(\sigma^{\mu}\bar{\lambda}_{\mu})_\alpha-(\sigma^{\nu}\bar{\sigma}^{\mu}\chi)_\alpha F_{\mu\nu}
-4i\kappa\chi\sigma_{\mu}\bar{\chi}\lambda^\mu_\alpha  \nonumber \\
 F_{\varpi \varpi } = & \hspace{-5pt} D^{2}-\frac{1}{2}F_{\mu\nu}F^{\mu\nu}-\frac{1}{2}iF_{\mu\nu}\tilde{F}^{\mu\nu}- 2i\chi\sigma^{\mu}\nabla_{\mu}\bar{\chi}
 -2\kappa D\chi\sigma^{\mu}\bar{\lambda}_{\mu}+4i\kappa\chi\sigma^{\nu}\bar{\lambda}^{\mu}F_{\mu\nu}  \nonumber \\
 & \hspace{-5pt} +i\kappa\lambda_{\rho}\sigma^{\nu}\bar{\sigma}^{\mu}\sigma^{\rho}\bar{\chi}F_{\mu\nu}
 +\frac{1}{2}\kappa^{2}\chi^{2}\bar{\lambda}_{\mu}\bar{\sigma}^{\mu}\sigma^{\nu}\bar{\lambda}_{\nu}
 -\frac{1}{2}\kappa^{2}\bar{\chi}^{2}\lambda_{\mu}\sigma^{\mu}\bar{\sigma}^{\nu}\lambda_{\nu} \\
 & \hspace{-5pt} -  4\kappa^{2}\lambda_{\nu}\sigma^{\nu}\bar{\lambda}_{\mu}\chi\sigma^{\mu}\bar{\chi}
 -4\kappa^{2}\chi\lambda_{\mu}\bar{\lambda}_{\nu}\bar{\sigma}^{\nu}\sigma^{\mu}\bar{\chi}
 -2\kappa^{2}\bar{\chi}^{2}\lambda_{\mu}\lambda^{\mu}-4\kappa^{2}\chi^{2}\bar{\lambda}_{\mu}\bar{\lambda}^{\mu}  \nonumber
\ea
For the canonical case $\mathscr{H}(\hat\Phi)=1$, one gets
\begin{eqnarray}
S_{\rm V}^{\rm NL} & = & \frac{1}{2}\int d^{4}x \det\emph{T} \left[D^{2}-i\chi\sigma^{\mu}\nabla_{\mu}\bar{\chi}+i\nabla_{\mu}\chi\sigma^{\mu}\bar{\chi}
-\frac{1}{2}F_{\mu\nu}F^{\mu\nu} \right. \\
 & &  +2\kappa\chi\sigma^{\nu}\bar{\lambda}^{\mu}\tilde{F}_{\mu\nu}+2\kappa\lambda^{\mu}\sigma^{\nu}\bar{\chi}\tilde{F}_{\mu\nu}
 +2i\kappa\chi\sigma^{\nu}\bar{\lambda}^{\mu}F_{\mu\nu}-2i\kappa\lambda^{\mu}\sigma^{\nu}\bar{\chi}F_{\mu\nu}\nonumber \\
 & & + 2\kappa^{2}\chi\sigma^{\mu}\bar{\lambda}_{\mu}\lambda_{\nu}\sigma^{\nu}\bar{\chi}-2\kappa^{2}\chi\lambda_{\mu}\bar{\lambda}_{\nu}\bar{\sigma}^{\nu}\sigma^{\mu}\bar{\chi}
 -2\kappa^{2}\bar{\chi}\bar{\lambda}_{\mu}\lambda_{\nu}\sigma^{\nu}\bar{\sigma}^{\mu}\chi\nonumber \\
 & &\left. - \frac{1}{2}\kappa^{2}\chi^{2}\bar{\lambda}_{\mu}\bar{\sigma}^{\mu}\sigma^{\nu}\bar{\lambda}_{\nu}
 -\frac{1}{2}\kappa^{2}\bar{\chi}^{2}\lambda_{\mu}\sigma^{\mu}\bar{\sigma}^{\nu}\lambda_{\nu}
 -2\kappa^{2}\bar{\chi}^{2}\lambda_{\mu}\lambda^{\mu}-2\kappa^{2}\chi^{2}\bar{\lambda}_{\mu}\bar{\lambda}^{\mu} \right]~\nonumber\label{eq:2}
\end{eqnarray}
Non-Abelian actions can be obtained from these by modifications as given in (\ref{non-abelian}).

For Abelian gauge theories, there could be the so-called Fayet-Iliopoulos terms\footnote
{There are strong arguments against the presence of Fayet-Iliopoulos terms,
in the context of supergravity \cite{Seiberg1,Dienes}. But they are included here for the sake of completeness.},
\be
S_{\rm FI} = 2 \xi_{\rm FI} \int d^4 x d^4 \theta \hat V(x, \theta, \bar \theta)
\ee
Converted into nonlinear realization by replacing $(x, \theta,\bar \theta)$ with $(z, \theta^{'},\bar \theta^{'})$, one has
\ba
S_{\rm FI}^{\rm NL} &=& 2 \xi_{\rm FI} \int d^4 x d^4 \theta \det \emph{T} \det \emph{M} V(x, \theta, \bar \theta) \nonumber \\
&=& \xi_{\rm FI} \int d^4 x \det \emph{T}
\left( D + \kappa \lambda_\mu \sigma^\mu \bar\chi +\kappa \chi\sigma^\nu \bar \lambda_\nu
-2 i \kappa^2 \epsilon^{\mu\nu\rho\sigma} \lambda_\mu \sigma_\sigma \bar\lambda_\nu v_\rho \right)
\ea
The last term in the parentheses seems to be non-covariant with respect to the residual gauge symmetry (\ref{r_gauge}).
It turns out that its infinitesimal gauge transformation times $\det\emph{T}$ is a total derivative \cite{IK3}.

The coupling between a chiral superfield and an Abelian gauge superfield is given by
\be
S_\mathcal{K}^{\rm V} =
\int d^4 x d^2\theta d^2\bar \theta  \Phi^\dagger e^{2gV} \Phi
\ee
which can be dealt with in the same manner as $S_\mathcal{K}$.
That is, to substitute $(x, \theta,\bar \theta)$ by either $(z, \theta^{'},\bar \theta^{'})$ or $(z_\pm, \theta^{'},\bar \theta^{'})$.
The former can be evaluated by the general procedure given in Appendix C.
The latter is simpler to use and one has
\be S_\mathcal{K}^{\rm V, NL} =
{1\over 2} \int d^4 x d^2\theta d^2\bar \theta \det \emph{T} \det \emph{M}_+
\left[ e^{-2i\theta \sigma^\mu \bar\theta \triangle_\mu^-} \varphi^\dagger(x,\bar\theta) \right] e^{2g U} \varphi(x,\theta) +h.c.
\ee
Explicitly, one has $S_\mathcal{K}^{\rm V, NL}$ as the real part of
\be \begin{array}{ll}
\int d^4 x \det \emph{T} \left[ \right.& \hspace{-10pt}
\mathscr{D}_\mu \mathscr{D}^\mu\phi^\dagger \phi + i\mathscr{D}_\mu \bar\psi \bar{\sigma}^\mu \psi+F^\dagger F 
+gD\phi^\dagger \phi+\sqrt2 ig\phi^\dagger\chi\psi-\sqrt2 i g\bar\psi\bar\chi\phi  \\
&\hspace{-10pt} + \kappa\left(-\sqrt2 \bar\lambda_\mu\bar\sigma^\nu\sigma^\mu \mathscr{D}_\nu \bar\psi\phi
+\sqrt2  \mathscr{D}_\nu\phi^\dagger \lambda_\mu\sigma^\nu\bar\sigma^\mu\psi
 \right.\\
&\hspace{12pt} \left.+\sqrt2\bar\psi\nabla^\mu\bar\lambda_\mu\phi +g\phi^\dagger\lambda_\mu\sigma^\mu\bar\chi\phi
+g\phi^\dagger\chi\sigma^\mu\bar\lambda_\mu\phi\right)\\
&\hspace{-10pt}  +2\kappa^2\left( \bar\psi \bar\lambda_\mu \lambda_\nu \sigma^\mu \bar\sigma^\nu \psi
-2i \lambda_\nu \sigma^\mu\bar\sigma^{\nu\rho} \bar\lambda_\rho  \mathscr{D}_\mu\phi^\dagger \phi\right)\\
&\hspace{-10pt}\left.
-4\sqrt2 i\kappa^3\bar\lambda_\mu  \bar\psi \lambda_\nu \sigma^\mu\bar\sigma^{\nu\rho} \bar\lambda_\rho \phi \right]
\end{array}
\ee
Here the gauge covariant derivatives are
\be
\mathscr{D}_\mu
\left( \begin{array}{l} \phi \\ \psi \end{array} \right)
 = (\nabla_\mu + i g v_\mu) \left( \begin{array}{l} \phi \\ \psi \end{array} \right)
\ee
and $\nabla_\mu + i g v_\mu = (T^{-1})_\mu^\nu (\partial_\nu + i g v^C_\nu)$.
If $\Phi$ is in the representation $t^a$ of a non-Abelian gauge group,
one needs to make replacements:
$v_\mu \rightarrow v_\mu^a t^a$, $\chi \rightarrow \chi^a t^a$, $D \rightarrow D^a t^a$.

Just as in the case of linear SUSY, both $F/D$ fields are quadratic in
$S_V^{\rm NL} +¡¡S_{\rm FI}^{\rm NL}+ S_\mathcal{K}^{\rm V, NL} + S_\mathcal{W}^{\rm NL} $
so they can be integrated out via equations of motion.
For the canonical Kahler potential and $\mathscr{H}=1$, one has,
\be\left\{\begin{array}{l}
D = - \xi_{\rm FI} - g  \phi^\dagger \phi, \ \ \ {\rm Abelian} \\
D^a = - g \phi^\dagger T^a \phi, \ \ \ \ \ \ {\rm non-Abelian}
\end{array} \right.
\ee
Both assume the same forms as those of linear theories and
$F$ is determined by (\ref{f_eom}).

\section{Conclusions}
We have in this paper revisited the nonlinear realization of spontaneously broken $N=1$ supersymmetry.
We have shown that the constrained superfield formalism as proposed in \cite{Seiberg}
can be reinterpreted in the language of standard realization of nonlinear supersymmetry via a new and simpler route.
We have presented explicit formulas of actions for all renormalizable theories with or without gauge interactions.
Particular attentions have been paid to the kinetic energies of chiral and vector superfields.
The nonlinear version of the Wess-Zumino gauge was discussed.
In addition, relations had been worked out for canonical gauge fields and those emerged naturally in the nonlinear reformulation.
A general procedure was also provided to deal with theories of arbitrary Kahler potentials.

In this reformulation, both $F/D$ fields are quadratic in the nonlinear action,
so they can be integrated out via equations of motion.
But all other component fields in $\Phi$ or $V$ are kept.
Since they all transform into themselves and are independent of one another,
any of them can be integrated out without breaking the nonlinear SUSY.
Whether and how to integrate out a component field are dynamical questions.
When some component fields have masses much higher than the energy scale of the concerned physical process,
they can be integrated out.
Ignoring quantum fluctuations,
these heavy fields can be expressed in terms of the light ones via equations of motion.
To include quantum fluctuations, matching and renormalization group running will be used,
which can only be carried when perturbation theory is applicable.
For strongly coupled systems, non-perturbative procedures need then to be developed.
Either way, heavy fields are effectively substituted by a set of high order operators constructed out of light fields.

Here one can make a connection between effective theories thus obtained with those constructed directly out of light fields.
In the latter cases, heavy fields are automatically set to zero,
with their effects represented by all possible high order operators permitted by the nonlinear SUSY and other symmetries of the theory of arbitrary coefficients.
Starting with a fundamental theory and reformulated into the nonlinear version,
all these coefficients can be determined, at least in principle, by integrating out heavy fields systematically.

\section*{Acknowledgement}
This work is supported in part by the National Science Foundation
of China (10425525, 10875103) and National Basic Research Program of China (2010CB833000).

\renewcommand{\theequation}{A.\arabic{equation}}
\setcounter{equation}{0}
\section*{Appendix A: Explicit expressions for $\det \emph{T}$ and $\det\emph{M}$}

Up to a multiplicative constant, $\det \emph{T} $ is the so-called Akulov-Volkov action for the nonlinear Goldstino field,
which is ubiquitous in nonlinear actions.
Explicitly,
\ba
\det \emph{T} &=& 1 - i\kappa^2 (\partial_\mu \lambda \sigma^\mu \bar \lambda - \lambda \sigma^\mu \partial_\mu \bar \lambda)  \\
&& + \kappa^4 \left[-i \epsilon^{\mu\nu\rho\gamma} \lambda \sigma_\rho \bar \lambda \partial_\mu \lambda \sigma_\gamma  \partial_\nu\bar \lambda
- \bar\lambda^2 \partial_\mu \lambda \sigma^{\mu\nu} \partial_\nu \lambda
- \lambda^2 \partial_\mu\bar \lambda \bar\sigma^{\mu\nu}  \partial_\nu \bar \lambda \right]\nonumber \\
&& - i \kappa^6 \lambda^2  \bar \lambda \left[ \bar\sigma^\rho \partial_\rho \lambda  \partial_\mu \bar\lambda \bar\sigma^{\mu\nu} \partial_\nu \bar\lambda
+  2 \bar\sigma^\nu \partial_\mu \lambda  \partial_\nu \bar \lambda \bar \sigma^{\rho\mu} \partial_\rho \bar\lambda
 \right]\nonumber \\
&&  - i\kappa^6 \bar\lambda^2 \lambda \left[ \sigma^\rho \partial_\rho  \bar\lambda \partial_\mu \lambda \sigma^{\mu\nu}  \partial_\nu  \lambda
+  2 \sigma^\nu \partial_\mu \bar\lambda  \partial_\nu  \lambda \sigma^{\rho\mu} \partial_\rho \lambda
\right]\nonumber \\
&& + {\kappa^8} \lambda^2  \bar \lambda^2
[\partial_\mu \bar \lambda \bar \sigma^{\mu\nu} \partial_\nu \bar\lambda         \partial_\rho \lambda  \sigma^{\rho\gamma} \partial_\gamma \lambda
 + \partial_\mu \bar \lambda \bar \sigma^{\nu\gamma} \partial_\rho \bar\lambda      \partial_\nu \lambda  \sigma^{\mu\rho} \partial_\gamma \lambda\nonumber \\
&& + 4 \partial_\mu \bar \lambda \bar \sigma^{\mu\rho} \partial_\nu \bar\lambda     \partial_\rho \lambda  \sigma^{\gamma\nu} \partial_\gamma \lambda]\nonumber
\ea
The combination $\det \emph{T} \det \emph{M}$ appears in the Jacobian when one changes integration variables from $(z, \theta^{'}, \bar\theta^{'})$ to $(x, \theta, \bar \theta)$.
One has
\ba
\det \emph{M} &=& 1 + i\kappa (\lambda_\mu \sigma^\mu \bar \theta - \theta \sigma^\mu \bar \lambda_\mu)  \\
&& + \kappa^2 \left[-i \epsilon^{\mu\nu\rho\gamma} \theta \sigma_\rho \bar \theta \lambda_\mu    \sigma_\gamma \bar \lambda_\nu
- \bar\theta^2 \lambda_\mu \sigma^{\mu\nu} \lambda_\nu
- \theta^2 \bar \lambda_\mu \bar\sigma^{\mu\nu}  \bar \lambda_\nu \right] \nonumber \\
&& + i \kappa^3 \theta^2  \bar \theta \left[ \bar\sigma^\rho \lambda_\rho  \bar\lambda_\mu \bar\sigma^{\mu\nu} \bar\lambda_\nu
+  2 \bar\sigma^\nu \lambda_\mu \bar \lambda_\nu \bar \sigma^{\rho\mu} \bar\lambda_\rho
 \right]\nonumber \\
&&  + i\kappa^3 \bar\theta^2 \theta \left[ \sigma^\rho \bar\lambda_\rho \lambda_\mu \sigma^{\mu\nu} \lambda_\nu
+  2 \sigma^\nu \bar\lambda_\mu \lambda_\nu \sigma^{\rho\mu} \lambda_\rho \right]\nonumber \\
&& + {\kappa^4} \theta^2  \bar \theta^2
[\bar \lambda_\mu \bar \sigma^{\mu\nu} \bar\lambda_\nu \lambda_\rho \sigma^{\rho\gamma} \lambda_\gamma
 + \bar \lambda_\mu \bar \sigma^{\nu\gamma} \bar\lambda_\rho  \lambda_\nu  \sigma^{\mu\rho} \lambda_\gamma
 + 4 \bar \lambda_\mu \bar \sigma^{\mu\rho} \bar\lambda_\nu    \lambda_\rho  \sigma^{\gamma\nu}\lambda_\gamma]\nonumber
\ea
Notice that $\det \emph{T} $ can be obtained from $\det \emph{M}$ by substitutions:
$\theta \rightarrow -\kappa \lambda(x)$ and $\bar\theta \rightarrow -\kappa \bar\lambda(x)$, $\nabla_\mu \rightarrow \partial_\mu$.

\renewcommand{\theequation}{B.\arabic{equation}}
\setcounter{equation}{0}
\section*{Appendix B: Taylor expansions of $\Phi(x, \theta, \bar\theta)$ and $\Phi^\dagger(x, \theta, \bar\theta)$}

Notice that for linear chiral superfields,
\be\left\{ \begin{array}{l}
\hat \Phi(x,\theta,\bar\theta)  =  \hat \phi(x) + i \theta \sigma^\mu\bar\theta \partial_\mu\hat \phi(x)
+ {1\over 4} \theta^2\bar\theta^2\partial^2\hat \phi(x) \\
\ \ \ \ \ \ \ \ \ \ \ \ \ \ \ \ + \sqrt{2} \theta\hat \psi(x) - {i\over\sqrt{2}} \theta^2 \partial_\mu\hat \psi(x) \sigma^\mu \bar\theta + \theta^2 \hat F(x) \\
\hat \Phi^\dagger(x,\theta,\bar\theta) = \hat  \phi^\dagger(x) - i \theta \sigma^\mu\bar\theta \partial_\mu\hat \phi^\dagger(x)
+ {1\over 4} \theta^2\bar\theta^2\partial^2\hat \phi^\dagger(x) \\
\ \ \ \ \ \ \ \ \ \ \ \ \ \ \ \ + \sqrt{2} \bar\theta \bar{\hat \psi}(x) + {i\over\sqrt{2}}\bar\theta^2  \theta\sigma^\mu \partial_\mu\bar{\hat \psi}(x)
+ \bar\theta^2 \hat F^\dagger(x)
\end{array} \right.
\ee
Taylor expansions of the nonlinear $\Phi(x,\theta,\bar\theta) $ and $\Phi^\dagger(x,\theta,\bar\theta)$
in terms of $\theta$ and $\bar\theta$ can be obtained from them by the following substitution rules:
\be \left\{ \begin{array}{l}
(\hat \phi, \hat \psi, \hat F) \rightarrow  (\phi, \psi, F), \ \ \
(\hat \phi^\dagger, \bar{\hat\psi}, \hat F^\dagger) \rightarrow  (\phi^\dagger, \bar\psi, F^\dagger), \\
\partial_\mu \hat\phi \rightarrow \phi_\mu,\ \ \ \
\partial_\mu \hat\psi  \rightarrow \psi_\mu + i \sqrt2\kappa \phi_\nu \sigma^\nu \bar\lambda_\mu  \\
\partial_\mu \hat\phi^\dagger \rightarrow \phi_\mu^\dagger,\ \ \
\partial_\mu \bar{\hat\psi} \rightarrow \bar\psi_\mu - i \sqrt2\kappa \lambda_\mu\sigma^\nu \phi_\nu^\dagger \\
\partial^2 \hat\phi \rightarrow \nabla^\mu \phi_\mu + \sqrt2\kappa \psi^\mu \lambda_\mu
+ 2 i\kappa^2 \lambda^\mu \sigma^\nu \bar \lambda_\mu \phi_\nu \\
\partial^2 \hat\phi^\dagger \rightarrow \nabla^\mu \phi_\mu^\dagger + \sqrt2\kappa \bar\psi^\mu \bar\lambda_\mu
- 2 i\kappa^2 \lambda^\mu \sigma^\nu \bar \lambda_\mu \phi_\nu^\dagger
\end{array} \right.\label{ksub}
\ee
where
\be\left\{ \begin{array}{l}
\phi_\mu = \nabla_\mu\phi + \sqrt2  \kappa \psi \lambda_\mu, \ \ \ \ \psi_\mu = \nabla_\mu\psi + \sqrt2\kappa F \lambda_\mu \\
\phi_\mu^\dagger = \nabla_\mu\phi^\dagger + \sqrt2 \kappa \bar\psi \bar\lambda_\mu, \ \ \ \bar\psi_\mu = \nabla_\mu\bar\psi + \sqrt2\kappa F^\dagger \bar\lambda_\mu
\end{array} \right.
\label{phipsimu}
\ee

\renewcommand{\theequation}{C.\arabic{equation}}
\setcounter{equation}{0}
\section*{Appendix C: General Kahler potential with or without gauge couplings}
Here we provide a general procedure to deal with theories of arbitrary Kahler potentials.
We will start with the linear theory and provide a set of substitution rules to get the nonlinear action.
To simplify presentation, we will abuse notations by using the same fields to denote both linear and nonlinear fields,
with explicit explanations to indicate their meaning whenever necessary.

In the linear theory, the Kahler potential is a vector superfield.
Without gauge couplings, it can be Taylor expanded as follows,
\ba
\mathcal{K}(\Phi^\dagger, \Phi) & = & \phi_\mathcal{K} + \theta \psi_\mathcal{K} + \bar\theta \bar\psi_\mathcal{K}
+ \theta \sigma^\mu \bar\theta v_{\mathcal{K}\mu} \\
&& + \theta^2 F_\mathcal{K} + \bar\theta^2 F_\mathcal{K}^\dagger
+ \theta^2 \bar\theta \bar\chi_\mathcal{K} + \bar\theta^2 \theta \chi_\mathcal{K} + \theta^2\bar\theta^2 D_\mathcal{K} \nonumber
\ea
where
\begin{eqnarray}
\phi_\mathcal{K} & = & \mathcal{K}(\phi^\dagger, \phi) \\
\psi_{\mathcal{K}\alpha} &=&   \sqrt{2} {\partial \mathcal{K} \over \partial \phi_i} \psi_{i\alpha}\\
F_\mathcal{K}  &=&  {\partial \mathcal{K} \over \partial \phi_i} F_i - {1\over 2}  {\partial^2 \mathcal{K} \over \partial \phi_i \partial \phi_j} \psi_i \psi_j   \\
v_{\mathcal{K}\mu} &=& i  {\partial \mathcal{K} \over \partial \phi_i} \partial_\mu \phi_i  -
i{\partial \mathcal{K} \over \partial \phi_i^\dagger}  \partial_\mu \phi^\dagger_i
+ {\partial^2 \mathcal{K} \over \partial \phi_i \partial \phi_j^\dagger} \bar\psi_{j} \bar\sigma_\mu \psi_{i}   \\
\chi_{\mathcal{K}\alpha} &=& {i\over \sqrt2}\left[ {\partial \mathcal{K} \over \partial \phi_i^\dagger} \sigma^\mu_{\alpha\dot\alpha} \partial_\mu \bar \psi_i^{\dot\alpha}
-  {\partial^2 \mathcal{K} \over \partial \phi_i^\dagger \partial\phi_j} \left(\sigma^\mu_{\alpha\dot\alpha} \bar\psi^{\dot\alpha}_i \partial_\mu \phi_j
+ 2i F_i^\dagger \psi_{j\alpha} \right) \right.\\
&&
\left. +  {\partial^2 \mathcal{K} \over \partial \phi_i^\dagger \partial\phi_j^\dagger} \sigma^\mu_{\alpha\dot\alpha} \partial_\mu \phi_i^\dagger \bar\psi_j^{\dot\alpha}
+ i  {\partial^3 \mathcal{K} \over \partial \phi_i^\dagger \partial\phi_j^\dagger \partial\phi_k} \bar\psi_i\bar\psi_j \psi_{k\alpha} \right]  \nonumber  \\
D_\mathcal{K} &=& \left[{1\over4} \partial_\mu \left( \partial^\mu \mathcal{K}
+2i  {\partial^2 \mathcal{K} \over \partial \phi_i^\dagger \partial\phi_j} \bar\psi_i \bar\sigma^\mu\psi_j \right)
 \right.   \\
&& + {\partial^2 \mathcal{K} \over \partial \phi_i^\dagger \partial\phi_j}
\left(  F_i^\dagger F_j - \partial_\mu\phi_i^\dagger\partial^\mu \phi_j
-i \bar\psi_i \bar\sigma^\mu\partial_\mu\psi_j \right)  \nonumber \\
&& - {1\over 2}  {\partial^3 \mathcal{K} \over \partial \phi_i \partial\phi_j \partial\phi_k^\dagger}  \psi_i \psi_j F_k^\dagger
- {1\over 2} {\partial^3 \mathcal{K} \over \partial \phi_i^\dagger \partial\phi_j^\dagger \partial\phi_k}
  \bar\psi_i \bar\psi_j F_k
  \nonumber \\&&
+ i {\partial^3 \mathcal{K} \over \partial \phi_i \partial\phi_j \partial\phi_k^\dagger} \partial_\mu \phi_i \psi_j \sigma^\mu \bar\psi_k
\left.  + {1\over 4}  {\partial^4 \mathcal{K} \over \partial \phi_i \partial\phi_j \partial\phi_k^\dagger \partial\phi_l^\dagger}
\psi_i \psi_j \bar\psi_k \bar\psi_l \right] \nonumber
\end{eqnarray}
When coupled to a gauge field, the Kahler potential becomes $\mathcal{K}(\Phi^\dagger, e^{2gV} \Phi)$,
which can be obtained from $\mathcal{K}(\Phi^\dagger,\Phi)$ by the following replacements:
\be\left\{ \begin{array}{l}
\partial_\mu\phi \rightarrow (\partial_\mu+ i g v_\mu)\phi,
\ \ \ \ \ \partial_\mu\psi \rightarrow (\partial_\mu+ i g v_\mu)\psi + {1\over \sqrt{2}}\bar \chi \bar\sigma_\mu \phi,\\
\partial_\mu\phi^\dagger \rightarrow (\partial_\mu - i g v_\mu)\phi^\dagger,
\ \ \  \partial_\mu\bar\psi \rightarrow (\partial_\mu - i g v_\mu)\bar\psi +{1\over \sqrt{2}}  \bar\sigma_\mu \chi \phi^\dagger,\\
\partial^\mu\partial_\mu\phi \rightarrow (\partial^\mu+ i g v^\mu) (\partial_\mu+ i g v_\mu)\phi  + 2 D \phi, \\
\partial^\mu\partial_\mu\phi^\dagger \rightarrow (\partial^\mu- i g v^\mu) (\partial_\mu - i g v_\mu)\phi^\dagger +2 D \phi^\dagger.
\end{array} \right.
\ee

To get nonlinear actions of theories without gauge couplings, we make the following replacements in the linear $\mathcal K(\Phi^\dagger, \Phi)$
by rules in (\ref{ksub}).
The nonlinear action is obtained by multiplying the expression thus obtained by $\det \emph{T} \det \emph {M}$ and integrated over $(x, \theta, \bar\theta)$.
When coupling to gauge fields, one needs further the replacements in the nonlinear actions thus obtained:
\be\left\{ \begin{array}{l}
\nabla_\mu\phi \rightarrow (\nabla_\mu+ i g v_\mu)\phi,
\ \ \ \ \ \nabla_\mu\psi \rightarrow (\nabla_\mu+ i g v_\mu)\psi + {1\over \sqrt{2}}\bar \chi \bar\sigma_\mu \phi,\\
\nabla_\mu\phi^\dagger \rightarrow (\nabla_\mu - i g v_\mu)\phi^\dagger,
\ \ \  \nabla_\mu\bar\psi \rightarrow (\nabla_\mu - i g v_\mu)\bar\psi +{1\over \sqrt{2}}  \bar\sigma_\mu \chi \phi^\dagger,\\
\nabla^\mu\nabla_\mu\phi \rightarrow (\nabla^\mu+ i g v^\mu) (\nabla_\mu+ i g v_\mu)\phi  + 2 D \phi, \\
\nabla^\mu\nabla_\mu\phi^\dagger \rightarrow (\nabla^\mu- i g v^\mu) (\nabla_\mu - i g v_\mu)\phi^\dagger +2 D \phi^\dagger.
\end{array} \right.
\ee
Note that both $F/D$ fields are quadratic in the nonlinear action.


\begin{thebibliography} {99}

\bibitem{WessBagger}
J. Wess and J. Bagger, ``Supersymmetry and Supergravity," Princeton Unversity Press (1992).

\bibitem{Weinberg}
S.~Weinberg,``The Quantum Theory of Fields,'' Vol II, Chapter 19,
Cambridge. Pr. (1996)

\bibitem{VA}
D.~V.~Volkov and V.~P.~Akulov, ``Is the Neutrino a Goldstone
Particle?,''  Phys.\ Lett.\  B {\bf 46}, 109 (1973).

\bibitem{Wess83}
 S.~Samuel and J.~Wess,
  ``A Superfield Formulation Of The Nonlinear Realization Of Supersymmetry And
  Its Coupling To Supergravity,''
  Nucl. Phys. B {\bf 221}, 153 (1983).

\bibitem{LLW}
H. Luo, M. Luo, and L. Wang,
``The Goldstino Field in Linear and Nonlinear Realizations of Supersymmetry,"
arXiv:0911.2836.

\bibitem{Seiberg}
Z. Komargodski, N. Seiberg, ``From Linear SUSY to Constrained
Superfields,'' arXiv:0907.2441.

\bibitem{LLZ}
H. Luo, M. Luo and S. Zheng, ``Constrained Superfields and Standard Realization of Nonlinear Supersymmetry",
e-Print: arXiv:0910.2110.

\bibitem{IK1}
  E.~A.~Ivanov and A.~A.~Kapustnikov,
  ``General Relationship Between Linear And Nonlinear Realizations Of
  Supersymmetry,''
  J. Phys. A  {\bf 11}, 2375 (1978).

\bibitem{IK2}
  E.~A.~Ivanov and A.~A.~Kapustnikov,
  ``Relation Between Linear And Nonlinear Realizations Of Supersymmetry,''
  JINR-E2-10765, Jun 1977.

\bibitem{IK3}
  E.~A.~Ivanov and A.~A.~Kapustnikov,
  ``The Nonlinear Realization Structure Of Models With Spontaneously Broken
  Supersymmetry,''
  J. Phys. G {\bf 8}, 167 (1982).

\bibitem{CL04}
 T.~E.~Clark and S.~T.~Love,
  ``Nonlinear realization of supersymmetry and superconformal symmetry,''
  Phys. Rev.  D {\bf 70}, 105011 (2004),
  [arXiv:hep-th/0404162].

\bibitem{CL96}
T.~E.~Clark and S.~T.~Love,
  ``Goldstino couplings to matter,''
  Phys.\ Rev.\  D {\bf 54}, 5723 (1996)
  [arXiv:hep-ph/9608243].

\bibitem{Klein}
M. Klein, ``Couplings in pseudosupersymmetry,'' Phys.~Rev.~D~{\bf
66}, 055009~(2002), [arXiv: hep-th/0205300].

\bibitem{Seiberg1}
Z. Komargodski and N. Seiberg, ``Comments on the Fayet-Iliopoulos Term in Field Theory and Supergravity",
JHEP 0906, 007 (2009).

\bibitem{Dienes}
K. R. Dienes and B. Thomas, ``A Proof of the Inconsistency of Fayet-Iliopoulos Terms in Supergravity Theories",
e-Print: arXiv:0911.0677.

\end{thebibliography}
\end{document}